\documentclass[12pt,a4j,aps,prc,superscriptaddress]{revtex4}
\usepackage[dvips]{graphicx}
\usepackage{amsmath,amssymb}
\usepackage{ulem}
\def\delsla{\!\not\!\partial}
\begin{document}
\title{
~\\[60pt]
Chiral phase transition and color superconductivity in an extended NJL
model with higher-order multi-quark interactions}

\author{Kouji Kashiwa}
\affiliation{Department of Physics, Graduate School of Sciences, Kyushu University,
             Fukuoka 812-8581, Japan}

\author{Hiroaki Kouno}
\affiliation{Department of Physics, Saga University,
             Saga 840-8502, Japan}

\author{Tomohiko Sakaguchi}
\affiliation{Department of Physics, Graduate School of Sciences, Kyushu University,
             Fukuoka 812-8581, Japan}

\author{Masayuki Matsuzaki}
\affiliation{Department of Physics, Fukuoka University of Education, 
             Munakata, Fukuoka 811-4192, Japan}

\author{Masanobu Yahiro}
\affiliation{Department of Physics, Graduate School of Sciences, Kyushu University,
             Fukuoka 812-8581, Japan}
\begin{abstract}
\begin{center}
{\normalsize Abstract}
\end{center}

\vspace{-4mm}

The chiral phase transition and color superconductivity in an 
extended NJL model with eight-quark interactions are studied.
The scalar-type nonlinear term hastens the chiral phase transition,
the scalar-vector mixing term suppresses effects of 
the vector-type linear term and 
the scalar-diquark mixing term makes the coexisting phase wider. 
\end{abstract}

\maketitle
%
Quantum Chromodynamics (QCD) has non-perturbative properties. 
First-principle lattice QCD simulations are useful to study 
thermal systems at zero or small density~\cite{Kog,ZF}.
At high density, however, lattice QCD is still not feasible 
due to the sign problem. Therefore, effective models are used 
in finite density region.
One of the models is the Nambu--Jona-Lasinio (NJL) model~\cite{NJ1}.  
 
This model has the mechanism of spontaneous chiral symmetry breaking,
but it has not the confinement mechanism.
However, this model has been widely 
used~\cite{Kle,HK1} with the mean field approximation (MFA), 
for example, for analyses of the critical endpoint 
of chiral phase transition~\cite{AY,Buballa,BR,Sca,Fuj,KKKN}. 

As for the NJL model, only a few studies were done so far 
on roles of higher-order multi-quark interactions~\cite{Osipov1,Osipov2}, 
except for the case of the six-quark interaction coming 
from the 't Hooft determinant interaction~\cite{tHooft}. 
The NJL model is an effective theory of QCD, so there is 
no reason, in principle, 
why higher-order multi-quark interactions are excluded. 

In this paper, we consider an extended NJL model that 
newly includes eight-quark interactions and analyze 
roles of such higher-order interactions on 
the chiral phase transition and color superconductivity.
It is well known that the original NJL model predicts 
a critical endpoint to appear at 
a lower temperature $(T)$ and a higher chemical potential ($\mu$) 
than the lattice QCD~\cite{ZF} and 
the QCD-like theory~\cite{KMT,HTF} do. 

As for the repulsive vector-type four-quark interaction 
$(\bar{q}\gamma^\mu q)^2$, 
it is well-known that it makes the chiral phase transition weaker 
in the low $T$ and high $\mu$ region 
and makes it a crossover when the vector type interaction is strong enough
~\cite{Buballa,KKKN}.
In this point of view, an $absence$ of 
the vector-type four-quark interaction 
may be preferable in the high density region.
On the contrary, a strong vector-type interaction is necessary
to reproduce the saturation property of nuclear matter
in the relativistic meson-nucleon theory~\cite{Walecka}.
Thus, it is expected that the vector-type interaction is sizable 
in the normal density region but suppressed in the higher density region.
In the relativistic meson-nucleon theory, it is known that 
nonlinear meson terms can suppress 
the effective coupling between mesons and 
nucleons in the higher density region.
Therefore, we consider higher-order multi-quark interactions in the NJL
model.

We start with the following chiral-invariant Lagrangian density 
with two flavor quarks 
\begin{eqnarray}
  {\cal L}&=&{\bar q}( i \delsla - m_0 )q
            + \Bigr[{g}_{2,0} \Bigl( 
               (\bar q q)^2 + (\bar q i \gamma_5  \vec{\tau} q)^2 \Bigl)
              +{g}_{4,0}\Bigl( 
               (\bar q q)^2 + ( \bar q i\gamma_5  \vec{\tau} q)^2 \Bigr)^2 
                                                          \nonumber \\
          &&  -{g_{0,2}}( \bar q \gamma^\mu q)^2
              -{g_{2,2}} \Bigl( (\bar q q)^2  
              +( \bar q i\gamma_5  \vec{\tau} q)^2 \Bigr)
	       (\bar q \gamma^\mu q)^2 \nonumber \\ 
          &&  +{d_{0,2}}(i{\bar q}^c \varepsilon \epsilon^b \gamma_5 q)
                  (i{\bar q} \varepsilon \epsilon^b \gamma_5 q^c)
	      +{d_{2,2}}(\bar q q)^2
                  (i{\bar q}^c \varepsilon \epsilon^b \gamma_5 q)
                  (i{\bar q} \varepsilon \epsilon^b \gamma_5 q^c)
+ \cdots \Bigr], \nonumber
\end{eqnarray}
where $q=q_{i \alpha}$ is the quark field, 
$q^c=C{\bar q^T}$ and ${\bar q}^c=q^T C$ are the charge-conjugation
spinor, $C=i\gamma^2\gamma^0$ is the charge-conjugation matrix,
$g_{i,j}$ and $~d_{n,m}$ are coupling constants, 
$m_0$ is the current quark mass, the Latin and Greek indices mean
the flavor and color,  ${\vec \tau}=(\tau^1~ \tau^2~ \tau^3)$ are Pauli
matrices and
$\varepsilon$ and $\epsilon^b$
are the antisymmetric in each of the flavor and color space.

In this paper, we use the MFA.
At the numerical calculation, we ignore higher-order terms 
represented by dots.
Furthermore, we disregard interactions including isovector-vector current 
not important in symmetric quark matter. 

We determine the parameters, $g_{2,0}$, $g_{4,0}$ and $\Lambda$, 
so as to reproduce the 
pion mass ($138~{\rm MeV}$), the sigma meson mass ($650~{\rm MeV}$) 
and the pion decay constant ($93.3~{\rm MeV}$).
The other
parameters $g_{0,2}$, $g_{2,2}$, $d_{0,2}$ 
and  $d_{2,2}$ are free parameters.
For the typical case, we assume that 
$g_{0,2}=Ag_{2,0},~g_{2,2}=B g_{2,0}/\sigma_0^2$,
$d_{0,2}=0.6g_{2,0},~d_{2,2}=Cd_{0,2}/\sigma_0^2$, 
$G_\sigma=2g_{2,0}+12g_{4,0}\sigma_0^2$, 
where $\sigma_0$ stands for the scalar density at $T=\mu=0$.
We take parameters 
as $(A,B,C)=(1.0,0,0)$, $(A,B,C)=(0.8,0.2,0)$ and
$(A,B,C)=(0,0,0.2)$. 
The current quark masses of up and down quarks are 
assumed to be $5.5~{\rm MeV}$.

Figure 1 shows the chiral phase transition line
in the $\mu$-$T$ plane~\cite{KKSMY}.\\

\begin{figure}[htpb]
\begin{center}
 \includegraphics[width=7.5cm]{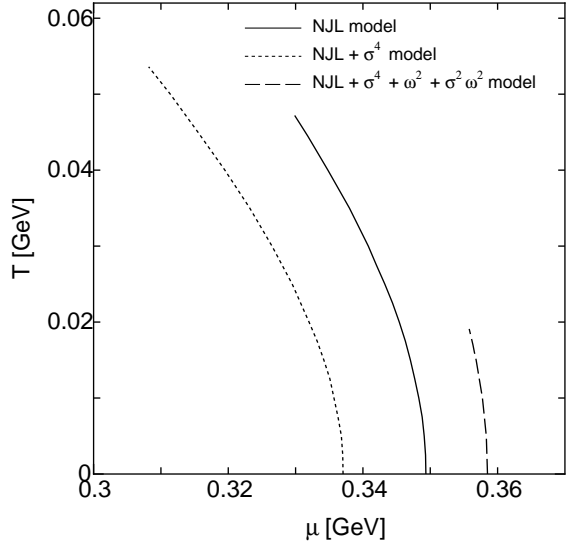}
\end{center}
\caption{Phase diagram in the $\mu$-$T$ plane. 
Each curve denotes the location of the first-order phase transition.}
\label{fig1}
\end{figure}
\begin{figure}[h]
\begin{center}
 \includegraphics[width=7.5cm]{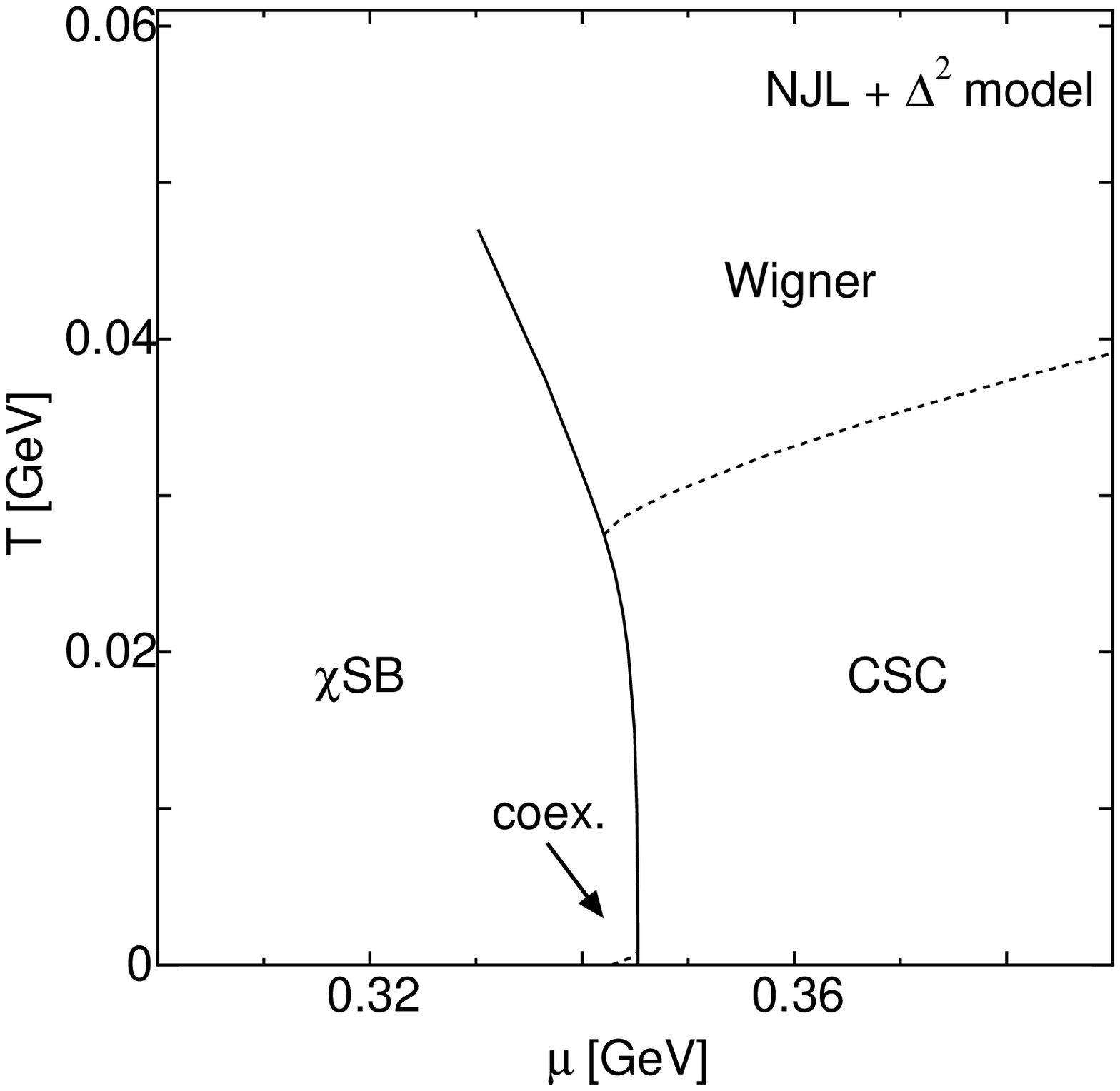} 
 \includegraphics[width=7.5cm]{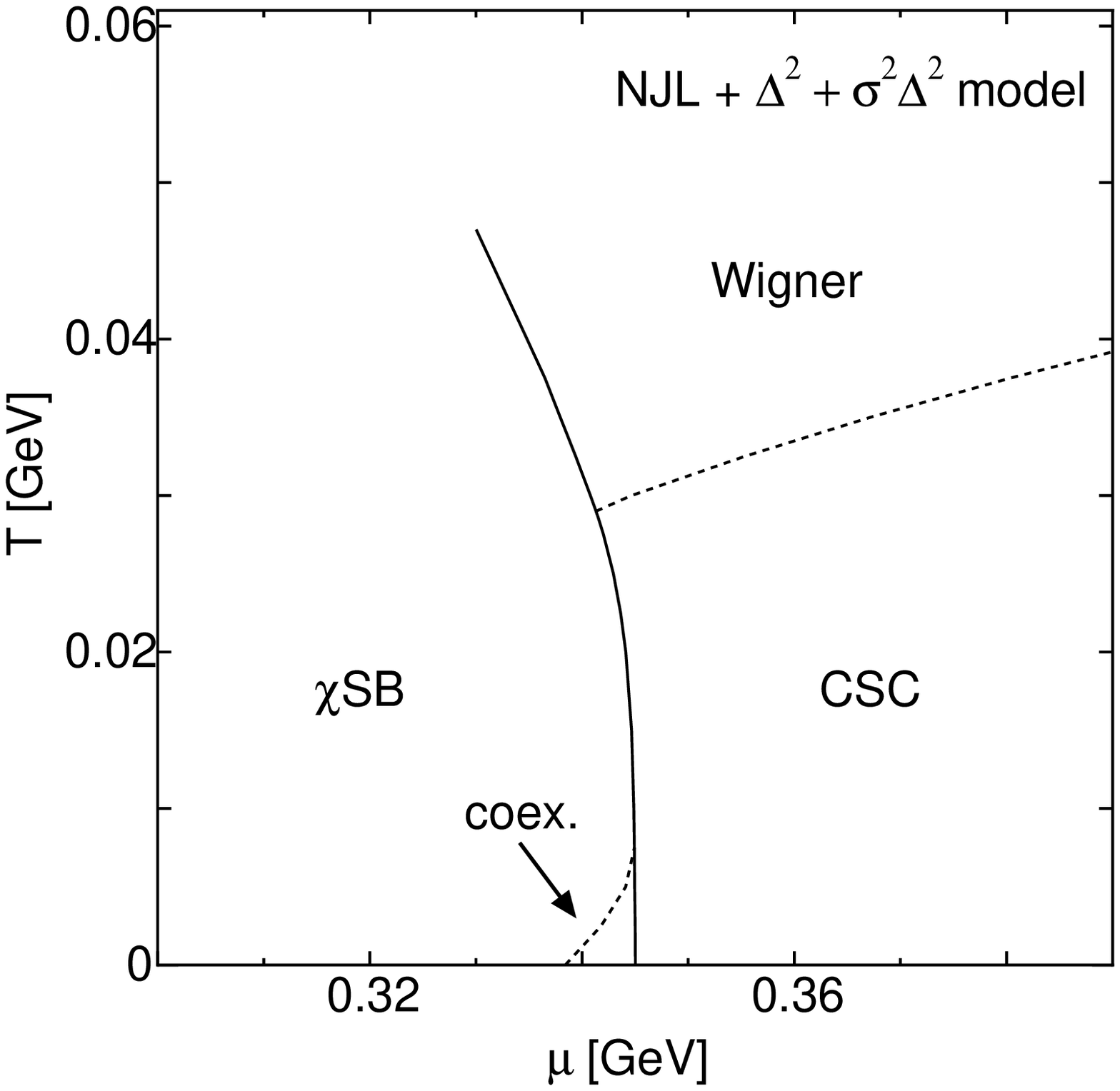} 
\end{center}
\caption{ Phase diagram in the $\mu$-$T$ plane. 
Solid lines represent the first order phase transition and
doted lines represent the second order phase transition.
The symbol ``$\chi$SB'' stands for the chiral symmetry broken phase,
``Wigner'' stands for the chiral symmetry restored phase,
``CSC'' stands for color superconductor phase and 
``coex.'' stands for the coexisting phase in which 
both the chiral condensation and the diquark
condensation have finite vacuum expectation values. }
\label{fig2}
\end{figure}

It is understood that the $\omega^2$ interaction tends 
to change the chiral phase transition 
from a first order to a crossover, but this effect is partially 
canceled out by the $\sigma^4$ and $\sigma^2\omega^2$ interaction. 
Consequently, as shown in Fig. 1, there exists a critical endpoint also 
in the NJL+$\sigma^4$+$\omega^2$+$\sigma^2\omega^2$ model.

Figure 2 shows 
the chiral phase transition line and color superconductivity
region in the $\mu$-$T$ plane.
Compare the left panel with the right one,
the $\sigma^2\Delta^2$ interaction makes the coexisting phase 
wider.

In conclusion, we have studied effects of eight-quark interactions
on the chiral phase transition and color superconductivity. 
The scalar-type nonlinear term $\sigma^4$ 
hastens the chiral phase transition and 
makes the critical endpoint move to a higher temperature and 
a lower chemical potential than original NJL model calculations. 
Both the scalar-type nonlinear term $\sigma^4 $ and 
the scalar-vector mixing term
$\sigma^2 \omega^2$
cancel effects of the vector-type linear term $\omega^2$.
The scalar-diquark mixing term $\sigma^2\Delta^2$
enlarges the coexisting region of 
the chiral condensate and the diquark condensate.
Therefore, higher-order multi-quark interactions are important 
for both the chiral phase transition and color superconductivity.


\end{document}